\documentclass[prl,preprint,tightenlines,%
superscriptaddress,showpacs,floatfix,nofootinbib
]{revtex4}
\usepackage{epsfig}

\begin{document}
\title{Mass and width of the Roper resonance using complex-mass renormalization}
\author{D.~Djukanovic}
\affiliation{Institut f\"ur Kernphysik, Johannes
Gutenberg-Universit\"at, D-55099 Mainz, Germany}
\author{J.~Gegelia}
\affiliation{Institut f\"ur Kernphysik, Johannes
Gutenberg-Universit\"at,  D-55099 Mainz, Germany}
\affiliation{High Energy Physics Institute, Tbilisi State University, Tbilisi,
Georgia}
\author{S.~Scherer}
\affiliation{Institut f\"ur Kernphysik, Johannes
Gutenberg-Universit\"at, D-55099 Mainz, Germany}
\date{\today}

\begin{abstract}

   The pole mass and the width of the Roper resonance are calculated as functions of
the pion mass in the framework of low-energy effective field theory of the strong interactions.
   We implement a systematic power-counting procedure by applying the complex-mass renormalization
scheme.

\end{abstract}

\maketitle

   At low energies, chiral perturbation theory \cite{Weinberg:1979kz,Gasser:1983yg}
provides a successful description of the Goldstone boson sector of QCD.
   Applying dimensional regularization in combination with the
modified minimal subtraction scheme leads to a straightforward
correspondence between the loop expansion and the chiral expansion
in terms of momenta and quark masses at a fixed ratio \cite{Gasser:1983yg}.
   In the momentum expansion, at each given order a systematic and
controllable improvement is possible.
   On the other hand, the issue of power counting in effective field theories with
heavy degrees of freedom turns out to be more complicated.
   For example, power counting is violated in baryon chiral perturbation theory,
if the dimensional regularization and the minimal subtraction scheme are used
\cite{Gasser:1987rb}.
   The problem may be solved by employing the heavy-baryon approach \cite{Jenkins:1990jv}
or by choosing a suitable renormalization scheme
\cite{Tang:1996ca,Becher:1999he,Gegelia:1999gf,Fuchs:2003qc}.
   Because the mass difference between the nucleon and the
$\Delta(1232)$ is small in comparison to the nucleon mass, the
$\Delta$ resonance can be consistently included in the framework of
effective field theory
\cite{Hemmert:1997ye,Pascalutsa:2002pi,Bernard:2003xf,Pascalutsa:2006up,Hacker:2005fh}.
   On the other hand, the inclusion of heavier baryon resonances such as the Roper resonance
is more complicated.
   We address the issue of power counting by using the complex-mass renormalization scheme
\cite{Stuart:1990,Denner:1999gp,Denner:2006ic,Djukanovic:2009zn},
which can be understood as an extension of the on-mass-shell
renormalization scheme to unstable particles.

\medskip

   We first specify those elements of the most general effective Lagrangian which are
relevant for the subsequent calculation of the pole of the Roper
propagator at order ${\cal O}(q^3)$:\footnote{Here, $q$ stands for
small parameters of the theory such as the pion mass.}
\begin{equation}
\mathcal{L} =  \mathcal{L}_{0}+ {\cal L}_\pi^{(2)}+ \mathcal{L}_{R} +\mathcal{L}_{N
R}+\mathcal{L}_{\Delta R}\,, \label{lagrFull}
\end{equation}
where $\mathcal{L}_{0}$ is given by
\begin{eqnarray}
\mathcal{L}_{0} & = & \bar{N}\,(i
D\hspace{-.65em}/\hspace{.1em}-m_{N 0})N + \bar{R}(i
D\hspace{-.65em}/\hspace{.1em}-m_{R 0}) R \nonumber\\
&&
 - \bar\Psi_\mu\xi^{\frac{3}{2}}
\biggl[(i {D\hspace{-.65em}/\hspace{.1em}}-m_{\Delta 0})\,g^{\mu\nu}
-i\,(\gamma^{\mu}D^{\nu}+\gamma^{\nu}D^{\mu})+i\,\gamma^{\mu}
{D\hspace{-.65em}/\hspace{.3em}}\gamma^{\nu} + m_{\Delta
0}\,\gamma^{\mu}\gamma^{\nu}\biggr]\xi^{\frac{3}{2}} \Psi_\nu .
\end{eqnarray}
   Here, $N$ and $R$ denote nucleon and Roper isospin doublets 
with bare masses $m_{N 0}$ and $m_{R 0}$, respectively.
   $\Psi_\nu$ are the vector-spinor isovector-isospinor
Rarita-Schwinger fields of the $\Delta$ resonance
\cite{Rarita:1941mf} with bare mass $m_{\Delta 0}$,
$\xi^{\frac{3}{2}}$ is the isospin-$3/2$ projector
(see Ref.~\cite{Hacker:2005fh} for more details).
   $D$ generically denotes covariant derivatives, which for the purposes of this work
may be replaced with ordinary partial derivatives.
   The lowest-order Goldstone boson Lagrangian reads
\begin{equation}
\label{l2}
{\cal L}_\pi^{(2)} =
\frac{F^2}{4}\mbox{Tr}\left(\partial_\mu U \partial^\mu
U^\dagger\right) +\frac{F^2 M^2}{4}\mbox{Tr} \left(
U^\dagger+ U \right)\,,
\end{equation}
where the pion fields are contained in the unimodular, unitary, $(2\times 2)$ matrix $U$.
   $F$ denotes the pion-decay constant in the chiral limit:
$F_\pi=F[1+{\cal O}(q^2)]=92.4$ MeV; $M$ is the pion mass at
leading order in the quark-mass expansion: $M^2=2 B\hat m$, where $B$
is related to the quark condensate $\langle \bar q q\rangle_0$ in the chiral
limit.

   The interaction terms $\mathcal{L}_{R}$, $\mathcal{L}_{NR}$, and $\mathcal{L}_{\Delta R}$
are constructed following Ref.~\cite{Borasoy:2006fk}.
   To leading order [${\cal O}(q)$], the pion-Roper coupling is given
by
\begin{equation}
{\cal L}_R^{(1)} = \frac{g_R}{2}\,\bar R \gamma^\mu\gamma_5 u_\mu
R\,, \label{Rpiint1}
\end{equation}
where $g_R$ is an unknown coupling constant and the pion fields are
contained in
\begin{equation}
u_\mu =i \left[u^\dagger \partial_\mu u -u \partial_\mu
u^\dagger\right], \label{umudef}
\end{equation}
where $u=\sqrt{U}$.
   In the present work we can neglect external sources except for the
quark mass term.
   The next-to-leading-order Roper Lagrangian relevant for our
calculation reads
\begin{equation}
{\cal L}_R^{(2)} = c_{1,0}^* \langle \chi_+\rangle\,\bar R\,R\,,
\label{Rpiint2}
\end{equation}
where $c_{1,0}^*$ is an unknown bare coupling constant and
$\chi_+ = M^2 (U+U^\dagger)$.
   The interaction between the nucleon and the Roper is given by
\begin{equation}
{\cal L}_{N R}^{(1)} = \frac{g_{N R}}{2}\,\bar R
\gamma^\mu\gamma_5 u_\mu N+ {\rm h.c.}\, \label{Rpiint11}
\end{equation}
with an unknown coupling constant $g_{N R}$.
   Finally, the leading-order interaction between the delta and the Roper
reads
\begin{equation}
{\cal L}_{\Delta R}^{(1)}= - g_{\Delta R} \,\bar{\Psi}_{\mu}
\,\xi^{\frac{3}{2}} \,(g^{\mu\nu}
+\tilde{z}\,\gamma^{\mu}\gamma^{\nu})\, u_{\nu}\, R + {\rm h.c.}\,,
\label{pND}
\end{equation}
where $g_{\Delta R}$ is an unknown coupling constant and, in
analogy with the nucleon case \cite{Hacker:2005fh}, we take the
''off-mass-shell parameter''  $\tilde z=-1$.

\medskip

   To renormalize the loop diagrams, we apply the complex-mass
renormalization of
Refs.~\cite{Stuart:1990,Denner:1999gp,Denner:2006ic}.
   For our case this means that we split the bare parameters of the Lagrangian in
renormalized parameters and counterterms and choose the
renormalized masses as the poles of the dressed propagators
in the chiral limit:
\begin{eqnarray}
m_{R 0} & = & z_\chi+\delta z_\chi\,,\nonumber \\
m_{N 0} & = & m+\delta m \,,\nonumber\\
m_{\Delta 0} & = &
z_{\Delta \chi} + \delta z_{\Delta \chi} \,,\nonumber\\
c_{1,0}^* & = & c_{1}^*+\delta c_{1}^*\,,\nonumber\\
&\cdots & \,, \label{barerensplit}
\end{eqnarray}
where $z_{\chi}$ is the complex pole of the Roper propagator in
the chiral limit, $m$ is the mass of the nucleon in the chiral limit,
and $z_{\Delta \chi}$ is the pole of the delta propagator in the
chiral limit.
   The ellipses stand for other parameters of the Lagrangian.
   We include the renormalized parameters $z_\chi$, $m$, and
$z_{\Delta \chi}$ in the propagators and treat the counterterms
perturbatively.

\medskip

   We organize our perturbative calculation by applying the standard
power counting of Refs.~\cite{Weinberg:1991um,Ecker:1995gg} to the
renormalized diagrams, i.e., an interaction vertex obtained from
an ${\cal O}(q^n)$ Lagrangian counts as order $q^n$, a pion
propagator as order $q^{-2}$, a nucleon propagator as order
$q^{-1}$, and the integration of a loop as order $q^4$.
   In addition, we assign the order $q^{-1}$ to the
$\Delta$ propagator and to the Roper propagator.
   Within the complex-mass renormalization scheme, such a power counting is
respected by the renormalized loop diagrams in the range of energies close to
the Roper mass.
   We implement this scheme by subtracting the loop diagrams at complex ''on-mass-shell''
points in the chiral limit.

\medskip

   The dressed propagator of the Roper is of the form
\begin{equation}
i S_R(p) = \frac{i}{p\hspace{-.45 em}/\hspace{.1em}-z_\chi
-\Sigma_R(p\hspace{-.45 em}/\hspace{.1em})}\,,\label{dressedDpr}
\end{equation}
where $-i\Sigma_R (p\hspace{-.45 em}/\hspace{.1em})$ denotes the sum of
one-particle-irreducible diagrams contributing to the Roper two-point
function.
   The dressed propagator $S_R$ has a complex pole which is obtained
by solving the equation
\begin{equation}
z - z_\chi -\Sigma_R(z)=0\,. \label{poleequation}
\end{equation}
   We define the pole mass and the width as the real part and $(-2)$ times
the imaginary part of the pole \cite{Djukanovic:2007bw}, respectively,
\begin{equation}
z = m_R -i\,\frac{\Gamma_R}{2} \,.\label{poleparameterized}
\end{equation}
   The mass and the width of an unstable particle defined through
the pole of the propagator are physical quantities.
   Such a definition guarantees that both are field-redefinition and gauge-parameter
independent \cite{'tHooft:1972ue,Lee:1972fj,Balian:1976vq}.

\medskip

\begin{figure}
\epsfig{file=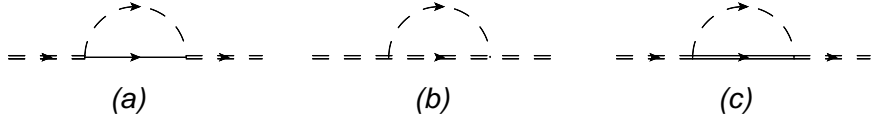,width=0.7\textwidth}
\caption[]{\label{DeltaMassInd:fig} One-loop self-energy diagrams
of the Roper. The dashed, solid, double-dashed, and double-solid lines
correspond to the pion, nucleon, Roper, and delta, respectively.}
\end{figure}

   To order ${\cal O}(q^3)$ the Roper self-energy consists of a
tree-order contribution
\begin{equation}
\Sigma_{\rm tree} =  4\,c_1^*M^2 \,,
\label{setree}
\end{equation}
and the loop diagrams shown in Fig.~\ref{DeltaMassInd:fig}.
   For the diagrams (a) and (b) of Fig.~\ref{DeltaMassInd:fig}
we obtain
\begin{eqnarray}
\Sigma_{(a)} & = & \frac{3 g_{N R}^2}{128 \pi^2 F^2}
\left[\hat{O}_1(m) A_0\left(m^2\right)
+\hat{O}_2(m)A_0\left(M^2\right)
+\hat{O}_3(m) B_0\left(p^2,m^2,M^2\right)\right], \label{sigma1}\\
\Sigma_{(b)} & = & \frac{3 g_R^2}{128 \pi^2 F^2 }
\left[\hat{O}_1(z_\chi) A_0\left(z_\chi^2\right)
+\hat{O}_2(z_\chi)A_0\left(M^2\right)
+\hat{O}_3(z_\chi)B_0\left(p^2,z_\chi^2,M^2\right)\right], \label{sigma2}
\end{eqnarray}
where
\begin{eqnarray*}
\hat{O}_1(x)&=&p\hspace{-.45 em}/\hspace{.1em}\,\left(1+ \frac{x^2}{p^2}\right)
+2x,\\
\hat{O}_2(x)&=&p\hspace{-.45 em}/\hspace{.1em}
\left(1-\frac{x^2}{p^2}\right),\\
\hat{O}_3(x)&=& p\hspace{-.45 em}/\hspace{.1em}\left[-p^2\left(1-\frac{x^2}{p^2}\right)^2
+M^2\left(1+\frac{x^2}{p^2}\right)\right]+2M^2 x.
\end{eqnarray*}
   Making use of dimensional regularization with $n$ the number of space-time
dimensions, the loop functions are defined as
\begin{eqnarray}
A_0\left(m^2\right) &=&  \frac{(2 \pi)^{4-n}}{i\,\pi^2}\,\int \frac{d^nk}{k^2-m^2+i\,0^+}\,, \nonumber\\
B_0\left(p^2,m_1^2,m_2^2\right) &=& \frac{(2 \pi)^{4-n}}{i\,\pi^2} \,
\int\frac{d^nk
}{\left[k^2-m_1^2+i\,0^+\right]\left[(p+k)^2-m_2^2+i\,0^+\right]}\,.\nonumber
\label{oneandtwoPF}
\end{eqnarray}
   Note the similarity of $\Sigma_{(b)}$ and $\Sigma_{(a)}$ resulting from the similarity
of the interaction Lagrangians $\mathcal{L}_{R}$ and $\mathcal{L}_{NR}$
of Eqs.~(\ref{Rpiint1}) and (\ref{Rpiint2}), respectively.
   However, $z_\chi$ is a complex quantity.
   The contribution of diagram (c) is given by
\begin{equation}
\label{sigma3}
\Sigma_{(c)} =  \frac{g_{\Delta  R}^2}{48 \pi^2 F^2 }
\left[\hat{O}_4
+\hat{O}_5 A_0\left(z_{\Delta \chi}^2\right)
+\hat{O}_6 A_0\left(M^2\right)
+\hat{O}_7B_0\left(p^2,z_{\Delta
\chi}^2,M^2\right)\right],
\end{equation}
where
\begin{eqnarray*}
\hat{O}_4&=&\frac{1}{6}\left[3 p\hspace{-.45 em}/\hspace{.1em}
z_{\Delta \chi}^2-12 p^2 z_{\Delta \chi}-4 p\hspace{-.45
em}/\hspace{.1em} p^2  + 4 p^2 \frac{p^2-3M^2}{z_{\Delta \chi}}+p\hspace{-.45 em}/\hspace{.1em}
\frac{2(p^2)^2-3 M^4-8 p^2 M^2}{z_{\Delta \chi}^2}\right],\\
\hat{O}_5&=&\frac{1}{p^2}
\left[p\hspace{-.45 em}/\hspace{.1em} z_{\Delta \chi}^2+2 p^2
z_{\Delta \chi} -p\hspace{-.45 em}/\hspace{.1em} \left(2
M^2+p^2\right) +2 p^2 \frac{p^2-M^2}{z_{\Delta \chi}}
+p\hspace{-.45 em}/\hspace{.1em}
\frac{\left(M^2-p^2\right)^2}{z_{\Delta\chi}^2}\right],\\
\hat{O}_6&=&-\frac{1}{p^2}\left[p\hspace{-.45 em}/\hspace{.1em} z_{\Delta \chi}^2+2 p^2
z_{\Delta \chi}-2 M^2 p\hspace{-.45 em}/\hspace{.1em} -2 p^2 \frac{M^2+p^2}{z_{\Delta \chi}}
+p\hspace{-.45em}/\hspace{.1em}\frac{M^4-3 p^2
M^2-(p^2)^2}{z_{\Delta\chi}^2}\right],\\
\hat{O}_7&=&-\frac{1}{p^2}\left[p\hspace{-.45 em}/\hspace{.1em} z_{\Delta \chi}^2+2 p^2
z_{\Delta \chi} +p\hspace{-.45 em}/\hspace{.1em}
\left(p^2-M^2\right)\right]
 \left[z_{\Delta \chi}^2-2 \left(M^2+p^2\right)
+\frac{\left(M^2-p^2\right)^2}{z_{\Delta\chi}^2}\right].
\end{eqnarray*}
   To implement the complex-mass renormalization scheme, in analogy
to Ref.~\cite{Fuchs:2003qc}, we expand the self-energy loop
diagrams in powers of $M$, $p\hspace{-.45 em}/\hspace{.1em}
-z_\chi$, and $p^2-z_\chi^2$, which all count as ${\cal O}(q)$.
   We subtract those terms which violate
the power counting, i.e., which are of order two or lower.
   The subtraction terms for the loop diagrams evaluated at
$p\hspace{-.45 em}/\hspace{.1em} =z_\chi$ are given by
\begin{eqnarray}
\Sigma_{(a)}^{\rm ST} & = &  -\frac{3 g_{N R}^2 (m+z_\chi)^2}{128 \pi^2
F^2 z_\chi} \biggl[(m-z_\chi)^2
B_0\left(z_\chi^2,0,m^2\right)-A_0\left(m^2\right)\biggr]\nonumber\\
&& + \frac{3 g_{N R}^2 (m+z_\chi) M^2}{64 \pi^2 F^2 z_\chi^3 }
\left[-2 m^3\,\ln\left(\frac{m}{\mu}\right) -i \pi m^3+z_\chi^2 m-32 \pi^2
z_\chi^3  \lambda \right.\nonumber\\
&&\left. +\left(m^3-z_\chi^3\right) \ln\left(\frac{z_\chi^2-m^2}{\mu^2}\right)
+i \pi
z_\chi^3
\right],\nonumber \label{STsigma1}\\
\Sigma_{(b)}^{\rm ST} & = &  \frac{3 g_R^2 z_\chi}{32 \pi^2 F^2}\
A_0\left(z_\chi^2\right) -\frac{3 g_R^2 z_\chi M^2}{32 \pi^2 F^2}
\left[32 \pi ^2 \lambda +2 \ln\left(
\frac{z_\chi}{\mu}\right)-1\right]\nonumber \label{STsigma2},\\
\Sigma_{(c)}^{\rm ST} & = & -\frac{g_{\Delta  R}^2}{288 F^2\pi^2
z_{\Delta \chi}^2 z_\chi} \left[
6 (z_{\Delta \chi}-z_\chi)^2 (z_{\Delta \chi}+z_\chi)^4B_0\left(z_\chi^2,0,z_{\Delta \chi}^2\right)
\right.\nonumber\\
&&+z_\chi^2
\bigl(-3 z_{\Delta \chi}^4
+12 z_\chi z_{\Delta \chi}^3
+4 z_\chi^2 z_{\Delta \chi}^2
-4 z_\chi^3 z_{\Delta \chi}
-2z_\chi^4\bigr)\nonumber\\
&&\left.
-6 \left(z_{\Delta \chi}^4+2 z_\chi z_{\Delta
\chi}^3-z_\chi^2 z_{\Delta \chi}^2+2 z_\chi^3 z_{\Delta
\chi}+z_\chi^4\right)
A_0\left(z_{\Delta \chi}^2\right)\right]\nonumber\\
&& + \frac{g_{\Delta  R}^2 M^2}{72 \pi^2 F^2 z_{\Delta \chi}^2 z_\chi^3}
\left[-6 i \pi z_{\Delta \chi}^6-6 (2 z_{\Delta \chi}+3
z_\chi) z_{\Delta \chi}^5\ln\left(\frac{z_{\Delta \chi}}{\mu}\right)
-9 i \pi z_\chi  z_{\Delta \chi}^5
+6 z_\chi^2 z_{\Delta \chi}^4\right. \nonumber\\
&& +9 z_\chi^3 z_{\Delta \chi}^3+3 z_\chi^4 z_{\Delta \chi}^2
-288 \pi ^2 \lambda z_\chi^5 z_{\Delta \chi}
+9 i  \pi z_\chi^5 z_{\Delta \chi}
+z_\chi^6
-192 \pi^2\lambda z_\chi^6 \nonumber\\
&& \left.+\left(6 z_{\Delta \chi}^6+9 z_\chi z_{\Delta \chi}^5-9
z_\chi^5 z_{\Delta \chi}-6 z_\chi^6\right)
\ln\left(\frac{z_\chi^2-z_{\Delta \chi}^2}{\mu^2}\right)+6 i \pi z_\chi^6
\right], \label{STsigma3}
\end{eqnarray}
where $\mu$ is the scale parameter of the dimensional
regularization and
\begin{equation}
\lambda =
\frac{1}{16\,\pi^2}\left\{\frac{1}{n-4}-\frac{1}{2}\,\left[\ln(4
\pi)+\Gamma'(1)+1\right]\right\}\,. \label{lambdadef}
\end{equation}

   The above expressions of Eq.~(\ref{STsigma3}) are exactly canceled
by counterterm contributions generated by $\delta z_\chi$ and
$\delta c_{1}^*$.
   Both of these quantities are complex
and none of them diverges in the limit of the nucleon mass and/or delta
mass approaching the Roper mass.

The pole of the Roper propagator to third order is given by the
expression
\begin{equation}
z = z_\chi -4\,c_1^*M^2 +
\left[\Sigma_{(a)}+\Sigma_{(b)}+\Sigma_{(c)}\right]_{p\hspace{-.3
em}/\hspace{.1em}=z_\chi}- \Sigma_{(a)}^{\rm ST}-\Sigma_{(b)}^{\rm
ST}-\Sigma_{(c)}^{\rm ST}\,. \label{pole}
\end{equation}
   The expansion of Eq.~(\ref{pole}) satisfies the power counting,
i.e.~is of ${\cal O}(q^3)$.
   In order to see this, we discuss the individual renormalized diagrams
of Fig.~\ref{DeltaMassInd:fig}.
   Let us start with diagram (b) (Roper resonance in the loop) and divide
the contribution by $M^3$.
   In the limit $M\to 0$ we find
$$
\lim_{M\to 0} \frac{1}{M^3}\left.\Sigma_{(b)}^{\rm ren}\right|_{p\hspace{-.3
em}/\hspace{.1em}=z_\chi}=-\frac{3 \,g_R^2}{32 \pi \,F^2}.
$$
   This is of the same type as the non-analytic contribution to the nucleon
mass if one replaces $g_R$ by $g_A$ \cite{Gasser:1987rb,Fuchs:2003qc}.

   Now let us turn to the delta contribution of diagram (c).
   Keeping the difference $z_\chi - z_{\Delta\chi}$ fixed and finite, the limit
$M\to 0$ is zero,
$$
\lim_{M\to 0} \frac{1}{M^3}\left.\Sigma_{(c)}^{\rm ren}\right|_{p\hspace{-.3
em}/\hspace{.1em}=z_\chi}=0.
$$
   If $z_\chi - z_{\Delta\chi}$ scales as $\alpha M$,
the limit $M\to 0$ is given by
\begin{displaymath}
\lim_{M\to 0} \frac{1}{M^3}\left.
\Sigma_{(c)}^{\rm ren}\right|_{p\hspace{-.3em}/\hspace{.1em}=z_\chi}
=\frac{g_{\Delta R}^2}{6 \pi^2 \,F^2}f(\alpha),
\end{displaymath}
where
\begin{displaymath}
f(\alpha)=4 i \pi
   \alpha ^3-6 i \pi  \alpha
   +\alpha +\left(6 \alpha -4 \alpha ^3\right) \ln (2 \alpha)
   +4
   \left(\alpha ^2-1\right)^{3/2} \ln \left(\alpha +\sqrt{\alpha
   ^2-1}\right)-4 i \pi  (\alpha ^2-1)^{3/2}.
\end{displaymath}
   Taking the limit $M\to 0$ after the limit $z_{\Delta \chi}\to z_\chi$ yields
$$
\lim_{M\to 0} \lim_{z_{\Delta \chi}\to z_\chi}\frac{1}{M^3}\left.
\Sigma_{(c)}^{\rm ren}\right|_{p\hspace{-.3em}/\hspace{.1em}=z_\chi}
=-\frac{g_{\Delta R}^2}{3 \pi \,F^2}.
$$
   In other words, whatever the counting for the mass difference, the contribution
of the renormalized diagram is of ${\cal O}(q^3)$.

   Finally, the nucleon contribution of diagram (a) results for
fixed and finite $z_\chi - m_N$ in
$$
\lim_{M\to 0} \frac{1}{M^3}\left.\Sigma_{(a)}^{\rm ren}\right|_{p\hspace{-.3
em}/\hspace{.1em}=z_\chi}=0.
$$
   If $z_\chi - m_N$ scales as $\beta M$,
the limit $M\to 0$ is given by
\begin{displaymath}
\lim_{M\to 0} \frac{1}{M^3}\left.
\Sigma_{(c)}^{\rm ren}\right|_{p\hspace{-.3em}/\hspace{.1em}=z_\chi}
=
\frac{3 \,g_{NR}^2}{64 \pi^2\,F^2 }f(\beta).
\end{displaymath}
  Taking the limit $M\to 0$ after the limit $m_N\to z_\chi$
 yields
$$
-\frac{3 \,g_{NR}^2}{32 \pi \,F^2 }.
$$

   The non-analytic terms of Eq.~(\ref{pole}) in powers of the
pion mass agree with the corresponding expression of
Ref.~\cite{Borasoy:2006fk}.
   On the other hand, our result is a closed expression and need not be expanded.
   Unlike the results of Ref.~\cite{Borasoy:2006fk} which were obtained
in the framework of the infrared renormalization scheme, our expression does
not diverge in the limit of the nucleon mass and/or delta mass approaching the Roper
mass.

\medskip

\begin{figure}
\epsfig{file=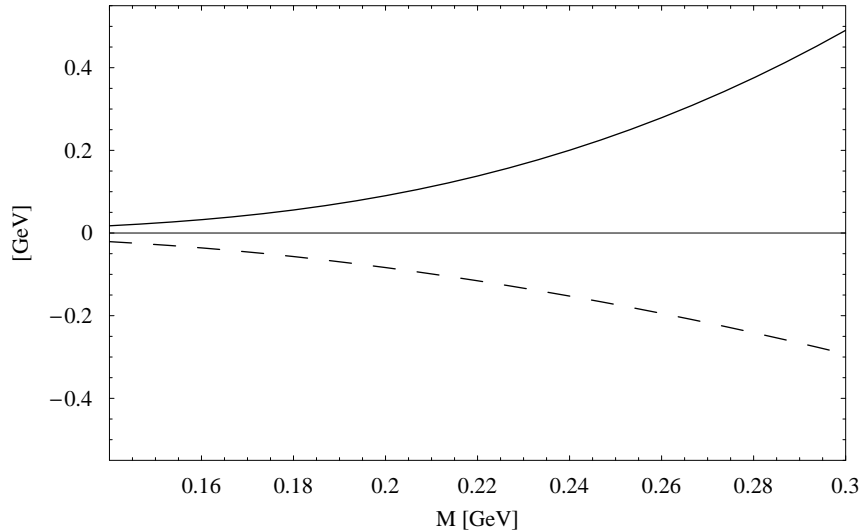,width=0.7\textwidth}
\caption[]{\label{MassWidth:fig}
Contributions of the renormalized loop diagrams
to the real (solid line) and imaginary (dashed line) parts of
the Roper pole as functions of the pion mass $M$.
}
\end{figure}

   To estimate the various contributions to the pole of the Roper
propagator, we substitute \cite{PDG} $F= 0.092\,\, {\rm GeV}$, $M = 0.140\,\, {\rm
GeV}$, $m=0.940\,\, {\rm GeV}$, $z_{\Delta \chi} = (1.210 - 0.100\,i/2)\,\,
{\rm GeV}$, $z_\chi = (1.365-0.190\,i/2)\,\, {\rm GeV}$, $\mu = 1\,\, {\rm
GeV}$, $g_R=1$, $g_{\Delta R} = 1$, $g_{NR}= 0.45$ \cite{Borasoy:2006fk} and obtain
\begin{equation}
z = \left[\left(1.365-\frac{i}{2}\, 0.190\right) -0.0784\,c_1^*
+ \left(0.0175- \frac{i}{2}\,0.042\right)\right]
\,{\rm GeV}\,. \label{polenumerical}
\end{equation}
   Figure \ref{MassWidth:fig} shows the contributions of the
renormalized loop diagrams to the real and imaginary parts of
the Roper pole as functions of the pion mass $M$.

\medskip

   To summarize, we have considered the chiral corrections to the
mass and the width of the Roper resonance in the framework of the
low-energy EFT of QCD.
   To obtain a systematic power counting for energies
around the mass of the Roper, we applied the complex-mass renormalization
scheme which is a generalization of the on-mass-shell renormalization for
the case of unstable particles.
   The mass and the width of the Roper in the chiral limit are considered
as input parameters of the approach.
   The chiral corrections have been calculated in a systematic way.

\medskip

\acknowledgments

   When calculating the loop diagrams we made use of the package
FeynCalc \cite{Mertig:1990an}. This work was supported by the
Deutsche Forschungsgemeinschaft (SFB 443).


\begin{thebibliography}{99}

\bibitem{Weinberg:1979kz}
S.~Weinberg,
Physica {\bf A96}, 327 (1979).

\bibitem{Gasser:1983yg}
  J.~Gasser and H.~Leutwyler,
  Annals Phys.\  {\bf 158}, 142 (1984).

\bibitem{Gasser:1987rb}
  J.~Gasser, M.~E.~Sainio, and A.~\v{S}varc,
  Nucl.\ Phys.\  {\bf B307}, 779 (1988).

\bibitem{Jenkins:1990jv}
  E.~E.~Jenkins and A.~V.~Manohar,
  Phys.\ Lett.\  B {\bf 255}, 558 (1991).

\bibitem{Tang:1996ca}
  H.~B.~Tang,
  arXiv:hep-ph/9607436.

\bibitem{Becher:1999he}
  T.~Becher and H.~Leutwyler,
  Eur.\ Phys.\ J.\  C {\bf 9}, 643 (1999).

\bibitem{Gegelia:1999gf}
  J.~Gegelia and G.~Japaridze,
  Phys.\ Rev.\  D {\bf 60}, 114038 (1999).

\bibitem{Fuchs:2003qc}
  T.~Fuchs, J.~Gegelia, G.~Japaridze, and S.~Scherer,
  Phys.\ Rev.\  D {\bf 68}, 056005 (2003).


\bibitem{Hemmert:1997ye}
  T.~R.~Hemmert, B.~R.~Holstein, and J.~Kambor,
  J.\ Phys.\ G {\bf 24}, 1831 (1998).

\bibitem{Pascalutsa:2002pi}
  V.~Pascalutsa and D.~R.~Phillips,
  Phys.\ Rev.\  C {\bf 67}, 055202 (2003).

\bibitem{Bernard:2003xf}
  V.~Bernard, T.~R.~Hemmert, and U.-G.~Mei{\ss}ner,
  Phys.\ Lett.\  B {\bf 565}, 137 (2003).

\bibitem{Pascalutsa:2006up}
  V.~Pascalutsa, M.~Vanderhaeghen, and S.~N.~Yang,
  Phys.\ Rept.\  {\bf 437}, 125 (2007).

\bibitem{Hacker:2005fh}
  C.~Hacker, N.~Wies, J.~Gegelia, and S.~Scherer,
  Phys.\ Rev.\  C {\bf 72}, 055203 (2005).

\bibitem{Stuart:1990}
R.~G.~Stuart, in ${\rm Z}^0$ {\it Physics}, ed. J. Tran Thanh Van
(Editions Frontieres, Gif-sur-Yvette, 1990), p.\ 41.

\bibitem{Denner:1999gp}
  A.~Denner, S.~Dittmaier, M.~Roth, and D.~Wackeroth,
  Nucl.\ Phys.\  {\bf B560}, 33 (1999).

\bibitem{Denner:2006ic}
  A.~Denner and S.~Dittmaier,
  Nucl.\ Phys.\ Proc.\ Suppl.\  {\bf 160}, 22 (2006).

\bibitem{Djukanovic:2009zn}
  D.~Djukanovic, J.~Gegelia, A.~Keller, and S.~Scherer,
  arXiv:0902.4347 [hep-ph].

\bibitem{Rarita:1941mf}
W.~Rarita and J.~S.~Schwinger,
Phys.\ Rev.\  {\bf 60}, 61 (1941).

\bibitem{Borasoy:2006fk}
  B.~Borasoy, P.~C.~Bruns, U.-G.~Mei{\ss}ner, and R.~Lewis,
  Phys.\ Lett.\  B {\bf 641}, 294 (2006).

\bibitem{Weinberg:1991um}
  S.~Weinberg,
  Nucl.\ Phys.\  B {\bf 363}, 3 (1991).

\bibitem{Ecker:1995gg}
  G.~Ecker,
  Prog.\ Part.\ Nucl.\ Phys.\  {\bf 35}, 1 (1995).

\bibitem{Djukanovic:2007bw}
  D.~Djukanovic, J.~Gegelia, and S.~Scherer,
  Phys.\ Rev.\  D {\bf 76}, 037501 (2007).

\bibitem{'tHooft:1972ue}
  G.~'t Hooft and M.~J.~G.~Veltman,
  Nucl.\ Phys.\  {\bf B50}, 318 (1972).

\bibitem{Lee:1972fj}
  B.~W.~Lee and J.~Zinn-Justin,
  Phys.\ Rev.\  D {\bf 5}, 3121 (1972).

\bibitem{Balian:1976vq}
D.~Gross, in {\it Methods in Field Theory},
edited by R.~Balian and J.~Zinn-Justin (North-Holland, Amsterdam, 1976).

\bibitem{PDG}
C.~Amsler {\it et al.} (Particle Data Group),
Phys.\ Lett.\ B {\bf 667}, 1 (2008).

\bibitem{Mertig:1990an}
  R.~Mertig, M.~Bohm, and A.~Denner,
  Comput.\ Phys.\ Commun.\  {\bf 64}, 345 (1991).

\end{thebibliography}
\end{document}